# Nonlocal and relativistic behavior form a mutual physical base: a toy model

Artur Szczepański[*]

Retired from the Institute of Fundamental Technological Research, Warszawa, Poland.

Classical objects have been excluded as subjects of the observed quantum properties, and the related problem of quantum objects' nature has been suspended since the early days of Quantum Theory. Recent experiments show that the problem could be reasonably revisited. The presented model indicates new issues which could result from exploring a specific kind of objects: topological defects in solids.

## Introduction

Foundations of Quantum Physics have ever been involved in metatheoretical queries. Historically, the related problems while significant in the interpretational debate, did not appear directly in the development of both the predictive power of the theory, and the applied research [1] until the advent of the contemporary Quantum Optics, and Quantum Communication. The questions, inter alia, linked to the status of theoretical quantum objects (e.g. wave functions), and to the status of quantum nonlocality, acquired an experimental and an applied meaning (See e.g. [2,3,4,5]). So some of the "quantum conundrum" problems weighting previously in the academic debate only, have emerged in applications, and in the new foundational issues.

Here I discuss a simple toy model of objects which are "excited states of the vacuum". Both the relativistic and nonclassical properties of the considered model objects result from a common single physical basis. This would be the prima facie most interesting feature of the model. However, it is important to emphasize that the model is evidently too simple to take it as an attempt to provide ready to use model-objects corresponding to real physical quantum-objects. The aim is rather to point at a category of physical objects of which both the nonclassical and the relativistic aspect are physically cohesive, so there would be no need to resort to the somewhat artificial concept of "peaceful coexistence" [1].

The introductory overview character makes me limit the discussion to a pre-formalized version easing to grasp the basic intuitions concerning the considered type of objects. Some meta-comments placing the model in a more general context are presented in the **Appendix**.

## The model

Consider topological point defects in a locally regular crystal matrix: vacancies (V) and interstitial atoms (A). V and A is created pairwise: an atom is hit out from a lattice node and placed in an interstitial position. The emptied node is the center of the vacancy. The dual process, that is the annihilation of the A-V pair consists in filling the vacancy by a lattice atom, and restoring the regular lattice structure. The pair-creation energy stored in the perturbed structure is released as lattice vibrations. The term "lattice atom" denotes in the present context an object the structure of which is irrelevant to the defect's properties.

---

[*]) Present address: artallszczep@gmail.com

[1]) The term has been coined by Abner Shimony [6] as describing the relative status of relativity and quantum nonlocality in the standard approach .

Both V and A are structural objects, (SO), formed by the displaced structure of the matrix. The defining state of A and V is characterized by the maximal symmetry of the nodes displacements, and shall be referred to as the static equilibrium state, (SES), of the defect. Note that the identity of A and V does not depend on the particular matrix's atoms involved in the displacement field of the actual defect.

The considered SO is different from a classical physical object: it should be taken neither for an object persisting in time (e.g. corpuscle), nor for a process (e.g. wave). See the **Appendix** for more details.

In the macro-description the lattice is modeled by a continuous medium. We are interested in media which can be treated as homogeneous, isotropic and incompressible.

Now, let us ask what could be the physics made by a "defect observer", (DO), that is an observer belonging to the "world of defects" [2].

"Making physics", grosso modo, is understood here in the standard way: formulating theories accounting for the results of observations. Hence the question: what observations are accessible to a DO, that is what kind of interactions feels a DO. The obvious answer is: interactions modifying the physical state of the DO. So the natural to the DO physics, which can be referred to as the "eigenphysics of defects", is based on interactions of structural perturbations of the matrix. Therefore a single DO placed in an otherwise unperturbed matrix is a free object since there is nothing which could act on his structure. He would than see an empty space, and the unperturbed matrix is to what a DO would refer to as the vacuum. This shows that he would have no direct experimental (observational) access to the underlying "world of the matrix's nodes-atoms", as expected.

So in the considered here model the notion of "physical matter" acquires a two-layer structure: the layer of "true matter" consisting of matrix atoms, and the layer of structural objects appearing as "matter" in the eigenphysics of defects.

The incompressibility of the matrix in the macro-description excludes the occurrence of longitudinal displacement waves in the medium, which opens the way for a **frictionless free propagation** process of both A and V [7]. Hence there may be matrixes allowing for eigenphysics in which a DO is unable to observe his free motion relative to the vacuum i.e. to the matrix (see, the **Appendix**): He can see free motions relative to other disturbed states of the matrix only, and in particular, to other DOs. In that case the DO could be led to assume the kinematical equivalence of reference frames linked to free-moving defects, and consequently to formulate a relativity principle in his eigenphysics.

The next step a DO could do is to conclude from the isotropy of the observed space, and from his relativity principle that the only admissible kinematical groups in his eigenphysics are the Galilei group and the Lorentz group [8]. But he would exclude the Galilei group since the relative velocity of defects, and consequently of reference systems in eigenphysics is bounded by the transverse (equivoluminal) wave velocity, $v_t$. Thus the DO would accept the Lorentz group with the invariant limiting velocity $v_t$ as the kinematical group in his eigenphysics.

---

[2]) The terms "defect observer" and "world of defects" should be taken for heuristic hints.

We turn back to the micro-level. Since the defect sees the perturbations of structure only, he has no direct observational access to the mechanics of the matrix nodes. This entails nonclassical effects in the eigenphysics of defects:

**1**. The set of time-evolving displacement vectors of the nodes forming the structure of a defect can be regarded as defining the time-evolution of the defect's state. The former are governed by the (invisible to the defects) mechanics of the lattice nodes, or in other words, by causes which do not belong to world of defects. So to a DO the observable time evolution contains an essentially indeterministic contribution, and this would lead to the appearance of an irreducible "true randomness" [9] in the eigenphysics of defects.

**2**. Consider a defects' state consisting of two parts which are spacelike separated in the eigenphysics. Suppose that a modification of the one part substate triggers a modification of the matrix's state resulting in the change of the second part's substate. Both the partial modifications are observable in the DO's eigenphysics. However their physical link, that is the interaction transfer in the underlying matrix, is a process occurring at the matrix level, and is not seen by a DO unless it is transmitted via matrix's displacement fields or other fields felt by defects. So such an action may be nonlocal to a DO because of lack of intermediary, and it may appear as superluminal, since $v_t$. is not the limiting velocity of the matrix atoms interactions.

Both ad. **1** and ad. **2** are examples of egzocosmic (relative to the "world of defects") actions the effects of which are observable to a DO. Such egzocosmic actions are not linked to "hidden parameters" in the sense of the historical debate about quantum foundations. The latter have been taken for endocosmic relative to the considered world of objects since the occurrence of egzocosmic physical effects is excluded in a single-layer material world. This would imply the nonexistence of egzocosmicity in the world of "true classical objects".

One more feature of the model: At the matrix level the properties of a V (A) in the SES state do not differ from the properties of another V (A) in its SES state, while the global state of the matrix is, obviously, different. This could be understood as the "structural indistinguishability" of defects at the matrix level, but to a DO, that is in the DO's eigenphysics, it would appear as "true indistinguishability".

**Conclusions**

We have considered a simple (actually the simplest I could imagine) model of a physical system with a two-layer "stratified matter". The basic layer, the "true matter" layer, consists of "atoms" forming a crystal-like structure (the matrix). The second layer in our model is the subsystem of structural point defects (vacancies and interstitial atoms) in the underlying matrix. Since the defects interact with structural perturbations of the matrix only, they do not "see" the atoms of the matrix, and the subsystem of defects can be regarded as a "quasi closed system". Consequently a "physicist" living in this "world of defects" would make physics limited by the accessibility of his observations to the "defects layer". At the macro-level, i.e. when the matrix is taken for a continuous (homogenous, isotropic, incompressible) medium, the defects behave classically. Their kinematical group would be the Lorentz one. On the other hand the defects' behavior is nonclassical at the micro-level: nonlocal, "truly random", etc. But both the relativistic and the nonclassical aspect have the common origin: the physics of the "true matter" layer.

Indicating the interest in exploring the idea of two-layer material world to the foundations of physics is the major aim of the present note.

**Appendix.**

The relation {Physical Theory ↔ Real Physical Objects} in classical physics ({CTh ↔ CO}), and in particular in Newtonian mechanics seems to be natural and well-understood in both the foundational and every-day research. Here I comment on the structure of classical physical objects, and the structure of the corresponding model objects to state expressis verbis some (often) silent assumptions needed to make the {CTh↔CO} so smoothly working. I than try to see how the corresponding assumptions function in our toy-model. Restating the latter proposal in the historical context of the XX century thirties would mean

> taking the Dirac sea for a solid medium, and treating matter and antimatter on the same footing. Since the antimatter has been formed out of holes in the sea, and since holes are "structural objects", matter should be taken for structural objects, too.

Let us treat {CTh↔CO} as a three-component structure consisting of:
1. The formalism of the theory, (F).
2. The theoretical model objects (MO).
3. The classical physical objects (CO).

In standard theories F is based on "sound mathematics", so F is the clearest part of the relation. The MOs are axiomatically endowed with a set of basic properties. At an appropriate level of generality the results of F applied to the MO-axioms establish new properties of MOs. The so deduced properties are interpreted as the predictions of the theory. The latter are tested via (experimental) observations, and the consistency of the observed CO's properties with the predicted MO's properties delimits the useful working range of {CTh↔CO}, that is (provisionally) establish the set physical objects to which properties the classical predictions apply.

The just outlined apparently obvious scheme works so naturally since the MOs are additionally endowed with basic features which make their general structure correspond to that of classical objects. To see it clearly we have to turn to the structure of classical objects. In the latter we may distinguish, according to Ingarden[3] [10], in addition to the just mentioned substructure of properties, two other substructures referred to as the categorial form and the mode of being.

The classical physical objects of interest have the categorial form of either an object persisting in time (e.g. corpuscles, bodies, media, etc.), or a process (e.g. motions of objects persisting in time). So in classical physics we have to do with either properties of objects persisting in time or properties of processes. The consistency of a MO's properties with the properties of the corresponding classical object can be achieved only, if the content of the MO is endowed with the categorial form of the corresponding classical object. For example a MO corresponding to a planet in Newton's theory is a mass point. Both have the categorial form of an object persisting in time. The orbital motion of a planet and the model-motion of the corresponding theoretical mass point both have the categorial form of a process.

---
[3]) For a concise outline and English terminology, see e.g. "Roman Ingarden" in the Stanford Encyclopedia of Philosophy.

Classical physical objects are supposed to be existentially independent of our conscious actions, and in particular of our cognitive actions aimed at those objects. This is the main existential requirement needed (in standard research) to interpret our experimental observations as observations of the real world.

Both the categorial form and the "existential autonomy" has been so natural and obvious to classical physicists that it has been assumed silently. Note however that the MOs themselves are theoretical constructs created by our conscious acts. On the other hand the content of their endowment i.e. the properties, the categorial form, and the existential mode once designed persists. Modification of the content requires the construction of a new MO. Otherwise the correspondence of the MOs and real objects would not be consistent.

Proceeding accordingly to the just outlined scheme has been regarded in the pre-quantum period as the major contribution to the progress of scientific knowledge concerning the physical reality. In practice the (classical) progress consists in accumulation of previously unknown properties of already known objects, or in revealing properties of new objects (persisting in time, or processes). The role of the remaining parts of the objects' structure (the silently assumed categorial form and mode of being) have been usually unnoticed.

Let us turn now to our toy-model. To "physicists of the matrix world" the defects' properties are just a subset of properties belonging to the matrix system. Therefore both their classical and nonclassical properties belong to well-defined subjects, i.e. to objects with easy to grasp and well-defined categorial form. The "existential autonomy" is there a natural feature. So model objects can be constructed, and the "matrix physicist" may proceed within the three-component scheme: {Theory – Model Objects – Matrix Objects}.

To a physicist living in the "defect world" (say, a DO of the main the text) observable properties would be consistent, if he assume them to belong to the structure-object layer in a two-layer-matter physical world. Accordingly he has to admit that he himself (say, his body) is a structural object. The classical physicist takes naturally himself (his body) for a "true matter" object. So the above stated basic assumption made by the "defect physicist" would be certainly a strong and bold assumption to him. Note moreover that apparently our metatheory does not provide tools to infer the categorial form of an object from its properties.

The structure-object-assumption would open new communication prospects to the "defect physicist": In his "defect world" there would occur exceptions to the non-signaling limitation (see, e.g. [11], and more recent papers by Peres), which would be of major interest to applications. The general importance of the assumption to the foundations and metatheory would consist in providing the answer to the question about the nature of the objects existing and forming the material world in which the "defect physicist" is living. Such an answer would enable him to work in the general framework of the Newtonian paradigm understood as in {CTh ↔ CO}.

Let me conclude with the following remark:
The Quantum Physicist of our world should admit the occurrence of rational reasons to look for the nature of quantum objects, that is to reconsider that fundamental question postponed in the early days of Quantum Theory: New issues could result in both application to quantum communication and foundations of physics.

## Acknowledgments

I am grateful to the participants of the ZTOC seminar at the Institute of Fundamental Technological Research (Warsaw), and of the seminar "Open Systems" at the Chair of Mathematical Methods of Physics of the Warsaw University for comments and suggestions.